\documentclass[journal,12pt,onecolumn,draftclsnofoot]{IEEEtran}
\usepackage{amsfonts}
\usepackage{graphicx}
\usepackage{booktabs}
\usepackage{algorithm}
\usepackage{algorithmic}
\usepackage{graphics}
\usepackage{subfigure}
\usepackage{multirow}
\usepackage{cite}
\usepackage{bigstrut,multirow,rotating}
\usepackage{amsmath,amsthm,amssymb}
\usepackage{array,diagbox,siunitx}
\usepackage{mathrsfs}
\usepackage{bm}
\usepackage{tikz}
\newcommand*{\circled}[1]{\lower.7ex\hbox{\tikz\draw (0pt, 0pt)%
    circle (.5em) node {\makebox[1em][c]{\small #1}};}}

\ifCLASSINFOpdf
\else
\fi
\begin{document}
\title{\LARGE{Self-information Domain-based Neural CSI Compression  with Feature Coupling}}

\author{Ziqing Yin,
        Renjie Xie,
        Wei Xu,~\IEEEmembership{Senior~Member,~IEEE},
        Zhaohui Yang,
        and Xiaohu You,~\IEEEmembership{Fellow,~IEEE} \vspace{-1cm}
        % <-this % stops a space
% \thanks{Copyright (c) 2015 IEEE. Personal use of this material is permitted. However, permission to use this material for any other purposes must be obtained from the IEEE by sending a request to pubs-permissions@ieee.org.}
% \thanks{This work was supported in part by the National Key Research and Development Program 2020YFB1806608;in part by the NSFC under grants 62022026 and 62211530108; in part by the Fundamental Research Funds for the Central Universities under grants 2242022K60002 and 2242023K5003.}
% \thanks{Ziqing Yin, Renjie Xie, Wei Xu, and Xiaohu You are with National Mobile Communications Research Laboratory, Southeast University, Nanjing 210096, China (email: zqyin@seu.edu.cn, renjie\_xie@seu.edu.cn, wxu@seu.edu.cn, xhyu@seu.edu.cn).}% <-this % stops a space
% \thanks{Zhaohui Yang is with the Zhejiang Laboratory, Hangzhou, Zhejiang 311121, China, also with the College of Information Science and Electronic Engineering, Zhejiang University, Hangzhou, Zhejiang 310027, China, and also with the Zhejiang Provincial Key Laboratory of Information Processing, Communication and Networking (IPCAN), Hangzhou, Zhejiang 310007, China(e-mail: yang\_zhaohui@zju.edu.cn).}% <-this % stops a space
% \thanks{Wei Xu is with National Mobile Communications Research Laboratory, Southeast University, Nanjing 210096, China(e-mail: wxu@seu.edu.cn).}% <-this % stops a space
}

\maketitle

\begin{abstract}
Deep learning (DL)-based channel state information (CSI) feedback methods compressed the CSI matrix by exploiting its delay and angle features straightforwardly, while the measure in terms  of information contained in the CSI matrix has rarely been considered. Based on this observation, we introduce self-information as an informative CSI representation from the perspective of information theory, which reflects the amount of information of the original CSI matrix in an explicit way. Then, a novel DL-based network is proposed for temporal CSI compression in the self-information domain, namely SD-CsiNet. The proposed SD-CsiNet projects the raw CSI onto a self-information matrix in the newly-defined self-information domain, extracts both temporal and spatial features of the self-information matrix, and then couples these two features for effective compression. Experimental results verify the effectiveness of the proposed SD-CsiNet by exploiting the self-information of CSI. Particularly for compression ratios $1/8$ and $1/16$, the SD-CsiNet respectively achieves $7.17$ dB and $3.68$ dB performance gains compared to state-of-the-art methods.    
\end{abstract}


% \fi
\IEEEpeerreviewmaketitle

\section{Introduction}

% \IEEEPARstart{M}{assive} multiple-input multiple-output (mMIMO) has become a key technology to improve the channel capacity and spectral efficiency for wireless communication networks. In order to fully exploit the advantages of mMIMO, the transmitter needs to obtain accurate channel state information (CSI) under the constraint of limited overhead and feedback bandwidth. In a frequency division duplexed (FDD) system, user equipments (UEs) estimate and compress the downlink CSI before feedback. Then the base station (BS) reconstructs the CSI matrix from the feedback information. However, the CSI matrix is large due to the growing number of antennas in mMIMO, which makes the CSI compression and feedback challenging. 

\IEEEPARstart{M}{assive} multiple-input multiple-output (mMIMO) has become a key technology for wireless communication networks. In order to fully exploit the advantages of mMIMO, the transmitter needs to obtain accurate channel state information (CSI). In order to reduce the overhead of feedback signaling,  researches have been devoted to developing efficient CSI compression methods \cite{xuedge2022}. For instance, codebooks have been utilized for CSI feedback in various commercial systems, e.g., LTE/LTE-A, IEEE $802.11$n/ac, and WiMAX. In \cite{shenchannel2018}, a codebook design based on angle-of-departure (AoD) was proposed for reducing the burden of CSI feedback. By utilizing slow-varying AoD information, the codebook was exploited to quantize the channel vector  within an angle coherence time period. However, both computational complexity and feedback overhead of the codebook-based CSI feedback methods increase sharply with the increase of the number of base station (BS) antennas \cite{luochannel2020}, which limits the practicability in mMIMO systems.

To overcome the challenges of codebook-based feedback methods in mMIMO systems, deep learning (DL) has been considered to apply for communication systems \cite{wang2023full}, especially for CSI feedback. By representing the CSI matrix as an image, a DL-based network, namely CsiNet, was originated in \cite{wendeep2018} by using the architecture of auto-encoder in DL. Existing DL-based studies mainly focused on improving the auto-encoder structure and network lightweight to enhance the feedback performance. In \cite{sunancinet2020}, an improved auto-encoder structure, called AnciNet, was proposed by considering that the obtained CSI was always noisy in practice, which achieved effective reconstruction for noisy inputs of CSI. An alternative DL network was proposed in \cite{songsaldr2021} to enhance the CSI feedback performance by introducing  self-attention learning and dense refine (SALDR). Further in \cite{lumimo2019}, a CSI compression network was proposed by exploiting the temporal correlation of wireless channels. In order to achieve a lightweight design of DL networks, a neural network with reduced size was proposed in \cite{jiclnet2021} for CSI compression by introducing a complex-valued input layer and self-attention mechanism. By considering shape and texture features of CSI images, a lightweight DL network, namely IdasNet, was proposed in \cite{yindeep2022} to perform accurate CSI recovery.

Essentially, from the perspective of information theory, the goal of CSI compression is to retain the amount of information of the channel matrix as much as possible under a limited feedback bandwidth. However, the original CSI image applied in \cite{wendeep2018,sunancinet2020,songsaldr2021,lumimo2019,jiclnet2021,yindeep2022} mostly considered the intrinsic features of the channels, i.e., delay and angle features. These DL-based methods compressed the CSI image by exploiting the intrinsic features rather than directly from the perspective of information contained in the CSI. Therefore, it is natural to establish a direct representation of the information of an mMIMO CSI matrix and then compress it based on this informative representation for better reconstruction at the BS.

% Essentially,  the object of CSI compression is to retain the information of the channel matrix as much as possible under a feedback limit for the CSI recovery at the BS. The more information contained in the feedback codeword, the more accurately the BS recovers the CSI. Most of these existing DL-based methods, however, compressed intrinsic channel parameters into a codeword straightforwardly from the perspective of image processing rather than from the perspective of the CSI information. In particular, the operation of a two-dimensional discrete Fourier transform (2D-DFT) that these DL-based methods applied highlights the sparsity of the transformed CSI image while not characterizing the informative features of the CSI.  Therefore, it is natural to establish a direct representation of the information in CSI and then compress the CSI based on this informative representation for better compression and reconstruction.  

In this paper, we introduce \emph{self-information} as the informative representation of CSI and compress the CSI in terms of the self-information. The definition of self-information is to measure and visually reflect the amount of block-wise information of the CSI matrix in an explicit way. We first project the original CSI matrix onto a self-information matrix in the newly-defined self-information domain. Then, a feature coupling encoder is proposed to extract both temporal and spatial features of the self-information representation. Note that the extracted temporal and spatial features are also correlated, thus we couple these two features for further compression. For recovery at the BS, a feature decoupling decoder is designed to recover the features and reconstruct the CSI matrix. Simulation results verify that the proposed method exhibits prominent performance gain compared to existing DL-based methods.

% The remainder of this paper is organized as follows. Section\text{ }\uppercase\expandafter{\romannumeral2} introduces the system model. Section \uppercase\expandafter{\romannumeral3} develops the proposed network and elaborates the design architecture. Section \uppercase\expandafter{\romannumeral4} presents the simulation results. The conclusion is drawn in Section \uppercase\expandafter{\romannumeral5}.

\section{System Model}
We consider the downlink of a frequency division duplexed (FDD) mMIMO system where the BS has $N_\text{t}$ antennas serving a user equipment (UE) with single antenna. Orthogonal frequency division multiplexing (OFDM) with $N_\text{s}$ subcarriers is employed. The received signal at the $n$th subcarrier is expressed as 
\begin{equation}
    y_n = \textbf{h}_n^H \textbf{v}_n x_n + g_n,
\end{equation}
\noindent where $\textbf{h}_n \in \mathbb{C}^{N_\text{t} \times 1}$, $\forall n \in \left \{1,\ldots, N_\text{s} \right\}$, denotes the channel vector at the $n$th subcarrier, $\textbf{v}_n \in \mathbb{C}^{N_\text{t} \times 1}$ is the precoding vector, $x_n \in \mathbb{C}$ is the transmitting signal, and $g_n \in \mathbb{C}$ represents the additive noise. The downlink channel matrix of all $N_\text{s}$ subcarriers is then denoted by $\textbf{H} = [\textbf{h}_1, \ldots, \textbf{h}_{N_\text{s}}]^H \in \mathbb{C}^{N_\text{s} \times N_\text{t}}$. The number of complex-valued feedback parameters of $\textbf{H}$ is therefore $N_\text{s} \times N_\text{t}$, which is enormous in mMIMO under a limited feedback bandwidth. In order to reduce the feedback overhead, we transform the channel matrix $\textbf{H}$ to the angular-delay domain, which explicitly presents possible channel sparsity \cite{sayeeddeconstructing2002}. By adopting the operation of two dimensional-discrete Fourier transform (2D-DFT), the angular-delay domain channel matrix is obtained as 
\begin{equation}
    \textbf{H}_\text{a} = \textbf{F}_\text{c} \textbf{H} \textbf{F}_\text{d},
\end{equation}
\noindent where $\textbf{F}_\text{c}$ and $\textbf{F}_\text{d}$ are the DFT matrices with proper dimensions. In the angular-delay domain, the channel $\textbf{H}_\text{a}$ contains useful information only in the first $N_\text{c}$ rows of multi-path delay with $N_\text{c}$ being the number of multiple paths, while the other rows are made up of near-zero values due to large propagation delays. Without loss of generality, we retain the first $N_\text{c}$ rows of $\textbf{H}_\text{a}$, denoted by $\textbf{H}_\text{f} \in \mathbb{C}^{N_\text{c} \times N_\text{t}}$, for the CSI compression.

Generally for a DL-based CSI compression method, due to that typical DL network training does not support complex-valued inputs\cite{sun2021a}, we separate the real and imaginary parts of $\textbf{H}_\text{f}$ as the input of the DL network, denoted by $\textbf{H}_\text{f} \in \mathbb{R}^{2 \times N_\text{c} \times N_\text{t}}$. Then an encoder deployed at the UE compresses the channel matrix $\textbf{H}_\text{f}$ to a specific codeword, denoted by $\textbf{c}$. The encoder network is represented by
\begin{equation}
    \textbf{c} = f_\text{EN}(\textbf{H}_\text{f}, \bf{\Theta_\text{E}}),
\end{equation}
\noindent where $f_\text{EN}(\cdot)$ denotes the compression function of the encoder and $\bf{\Theta_\text{E}}$ is the set of the training parameters. At the other side, a decoder network deployed at BS recovers the channel matrix based on the received codeword $\textbf{c}$, which is represented by
\begin{equation}
    \widehat{\textbf{H}}_\text{f} = f_\text{DE}(\textbf{c}, \bf{\Theta_\text{D}}),
\end{equation}
\begin{figure*}[t]
\centering
\includegraphics[scale=0.175]{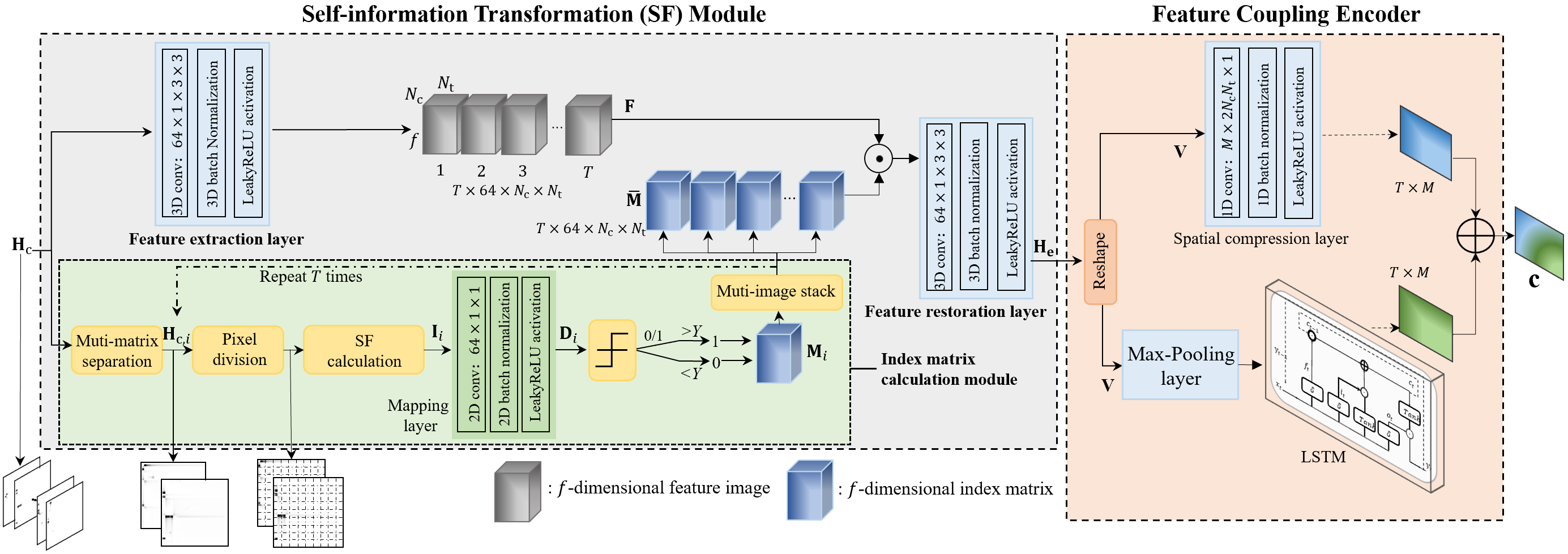}
\vspace{-0.6em}
\caption{Structure of the SF module and the feature coupling encoder at the UE.}
\label{fig:label2}
\end{figure*}
\noindent where $f_\text{DE}(\cdot)$ is the decompression function of the decoder and $\bf{\Theta_\text{D}}$ is the set of corresponding training parameters. The desired channel matrix, denoted by $\widehat{\textbf{H}}$, can be finally acquired by adopting an inverse 2D-DFT to $\widehat{\textbf{H}}_\text{f}$ in (4).

\section{Neural CSI Compression with Feature Coupling}
In this section, we elaborate the architecture of the proposed neural CSI compression network with temporal and spatial features under the self-information domain, referred to as SD-CsiNet. The proposed network consists of a self-information transformation (SF) module, an encoder with feature coupling, and a corresponding decoder with feature decoupling.  Without loss of generality, we  assume that there exists $T$ temporal CSI matrices within a coherent time. Therefore, it is necessary to introduce an additional dimension  upon $\textbf{H}_\text{f}$ to represent the number of temporal CSI matrices in a coherent time, which is denoted by $\textbf{H}_\text{c} \in \mathbb{R}^{T \times 2 \times N_\text{c} \times N_\text{t}}$. Then the recovered CSI image at BS is denoted by $\widehat{\textbf{H}}_\text{c} \in \mathbb{R}^{T \times 2 \times N_\text{c} \times N_\text{t}}$. Specifically, the SF module transforms the CSI matrix from the angular-delay domain to the newly-defined self-information domain. The transformed matrix in the self-information domain successfully highlights the informative characteristic of the CSI. Subsequently, the feature coupling encoder extracts both temporal and spatial features of the transformed matrix and  couples them into a codeword \textbf{c}. The decoder at the BS recovers the CSI matrix. The architecture details of each module in SD-CsiNet are elaborated in the following.

\subsection{The SF Module}
To better reflect the amount of information contained in the CSI matrix, the SF module transforms the CSI matrix from the angular-delay domain to the self-information domain. A concrete illustrative description of the SF module is depicted in Fig. 1. The module consists of a feature extraction layer, an index matrix calculation module, and a feature restoration layer. Note that the input of the SF module is a sequence of CSI matrices $\textbf{H}_\text{c}$. The output of the SF module is the self-information matrix of the CSI, denoted by $\textbf{H}_\text{e} \in \mathbb{R}^{T \times 2 \times N_\text{c} \times N_\text{t}}$.

The upper part in Fig. 1 is the feature extraction layer, which transforms $\textbf{H}_\text{c}$  to 64 feature maps, denoted by $\textbf{F} \in \mathbb{R}^{T \times 64 \times N_\text{c} \times N_\text{t}}$, where each feature map characterizes a specific feature of $\textbf{H}_\text{c}$. Different from existing DL-based methods, e.g., \cite{wendeep2018,sunancinet2020,songsaldr2021,lumimo2019,jiclnet2021,yindeep2022}, we adopt a three-dimensional (3D) convolutional layer with a filter size of $64 \times 1 \times 3 \times 3$, rather than  conventional 2D convolutional layers, because $\textbf{H}_\text{c}$ contains the time dimension. Here the dimension of $64$ denotes the number of feature maps and $1 \times 3 \times 3$ is the kernel size of the 3D convolutional layer.

The lower part is the index matrix calculation module, which is an essential component network for transforming $\textbf{H}_\text{c}$ to the self-information domain. It analyzes the pixel-wise entropy of the CSI matrix, calculates the self-information of CSI matrix, $\textbf{H}_\text{c}$, in an pixel-wise manner and outputs an index matrix. The index matrix rules out  pixels with marginal information in the feature matrices while retains pixels with large entropy in terms of the self-information. In the index matrix calculation module, we first separate $\textbf{H}_\text{c}$ into $T$ CSI matrices $\textbf{H}_{\text{c},i} \in \mathbb{R}^{2 \times N_\text{c} \times N_\text{t}}$ for $i \in \left \{1,2, \ldots, T \right \}$. Then we divide each of $\mathcal{R}(\textbf{H}_{\text{c},i})$ and  $\mathcal{I}(\textbf{H}_{\text{c},i})$ into $N_\text{c} \times N_\text{t}$ pixels by using a $1 \times 1$ dividing grid. Each pixel is denoted by $\text{p}_j \in \mathbb{R}^{ 1 \times 1}$ for $j \in \left \{1,2,\ldots,N_\text{c} N_\text{t} \right \}$.  According to the definition of information theory, the probability needs to be acquired before calculating the self-information of $\text{p}_j$. This probability relates to the surrounding pixels of $\text{p}_j$. Intuitively, if $\text{p}_j$ is significantly different from surrounding pixels, it contains more essential information for CSI and corresponds to a low probability. Thus by checking on the surrounding pixels and inspired by \cite{shiinformative2020}, the self-information of $\text{p}_j$ is calculated as
\begin{equation}
   \widehat{I}_j \!=\! - \text{log}_2 \frac{1}{9} \sum_{r=1}^9 \frac{1}{\sqrt{2 \pi }} \text{exp} \left (- \left \| \text{p}_j-\text{p}'_{j,r}\right \|_2^2/2 \right ),
\end{equation}
\noindent where $\widehat{I}_j$ denotes an estimate of self-information of $\text{p}_j$, and $\text{p}'_{j,r}$ denotes the $r$th surrounding pixel of $\text{p}_j$. Note that for a specific $\text{p}_j$, considering the balance of the running time and the network performance, we pick out $9$  pixels that are closet to the center pixel $\text{p}_j$, including itself, to calculate the self-information. That is, we have $r \in \left \{1,2, \ldots, 9 \right \}$. Given the definition of self-information in (5), we now obtain a self-information matrix, denoted by $\textbf{I}_i \in \mathbb{R}^{2 \times N_\text{c} \times N_\text{t}}$, and the $(2,m,n)$th element of $\textbf{I}_i$ is $\widehat{I}_j$. 

After the SF calculation, it follows a mapping layer in Fig. 1. This layer maps the self-information matrix $\textbf{I}_i$ to an informative feature matrix, denoted by $\textbf{D}_i \in \mathbb{R}^{64 \times N_\text{c} \times N_\text{t}}$. Considering there is no time dimension in $\textbf{I}_i$, the mapping layer adopts a 2D convolution with a filter size of $64 \times 3 \times 3$. Note that this mapping layer has non-trainable filter parameters and thus it is not involved for gradient update in network training. Note that the self-information $\widehat{I}_j$ is a numerical representation of the information contained in $\text{p}_j$. The more information that $\text{p}_j$ contains, the larger $\widehat{I}_j$ is. Accordingly, we set a self-information threshold, denoted by $Y$, to rule out pixels with smaller self-information values in $\textbf{D}_i$ to ensure that only the most informative pixels in the CSI matrix retain in the  compression. In particular, the elements in $\textbf{D}_i$ with their self-information smaller than $Y$ are set to zeros, and the other elements are all ones. Mathematically, this operation is expressed as
\begin{eqnarray}
m_j = 
\left\{
\begin{array}{lll}
1, \ \ {\rm{if}} \ d_j \geq Y \\
0, \ \ \text{otherwise},
\end{array}
\right.
\end{eqnarray}
\noindent where $d_j$ denotes the $j$th element of $\textbf{D}_i$ and $m_j$ denotes the position of $d_j$ in $\textbf{D}_i$. The corresponding index matrix $\textbf{M}_i \in \mathbb{R}^{64 \times N_\text{c} \times N_\text{t}}$ contains all these $\left \{m_j \right \}$'s $0$'s and $1$'s. Since the index matrix calculation module only calculates the index matrix of a single CSI matrix $\textbf{H}_{\text{c},i}$ at each time, this module runs $T$ times to get all the index matrices of $\textbf{H}_\text{c}$. The running times therefore corresponds to the first dimension of $\textbf{H}_\text{c}$. We then stack all the index matrices in $\bar{\textbf{M}} \in \mathbb{R}^{T \times 64 \times N_\text{c} \times N_\text{t}}$ as follows
\begin{equation}
    \bar{\textbf{M}} = [\textbf{M}_1,\textbf{M}_2,\cdot \cdot \cdot,\textbf{M}_T].
\end{equation}
After acquiring $\bar{\textbf{M}}$, the SF module uses operator $\textbf{F} \odot \bar{\textbf{M}}$ for removing the elements with trivial information in $\textbf{F}$, where $\odot$ denotes the Hadamard product. The last part in the SF module, i.e., the feature restoration layer with the filter size of $2 \times 1 \times 3 \times 3$, completes the transformation from $\textbf{F}$ to  $\textbf{H}_\text{e}$ in the self-information domain. The output of SF module highlights the informative characteristic of $\textbf{H}_\text{c}$ compared to the original CSI matrix, which is conductive for extracting  spatial and temporal features in the subsequent encoder network.

\begin{figure*}[t]
\centering
\includegraphics[scale=0.22]{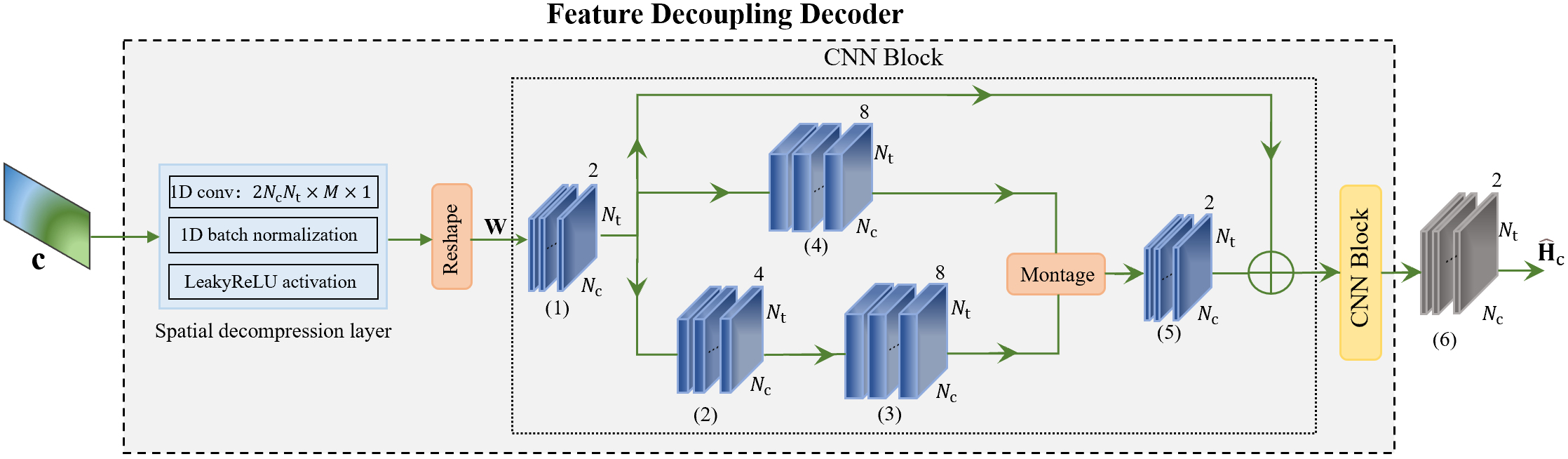}
\vspace{-0.7em}
\caption{The detailed structure of feature decoupling decoder at the BS.}
\label{fig:label3}
\end{figure*}

\subsection{Feature Coupling Encoder}
Given the self-information matrix $\textbf{H}_\text{e}$ of the CSI, the feature coupling encoder extracts both temporal and spatial features of $\textbf{H}_\text{e}$ and couples them to a codeword. This encoder consists of a spatial compression layer, a max-pooling layer, and a long short-term memory (LSTM) module as shown in the right part of Fig. 1. It outputs the feedback codeword $\textbf{c}$.

The self-information matrix $\textbf{H}_\text{e}$ is first reshaped into a matrix $\textbf{V} \in \mathbb{R}^{T \times 2 N_\text{c} N_\text{t}}$. The spatial compression layer adopts a 1D convolution with filter size of $M \times 2 N_\text{c} N_\text{t} \times 1$, rather than a popular fully-connected (FC) layer, to extract the spatial feature of $\textbf{V}$. Note that $M$ is the dimension of the feedback codeword, determined by $M = \sigma \times 2 N_\text{c} N_\text{t}$, where $\sigma$ is the compression ratio and $2 N_\text{c} N_\text{t} \times 1$ is the kernel size of the 1D convolutional layer. On the other hand, the max-pooling layer is used to reduce the dimension of $\textbf{V}$ from $T \times 2 N_\text{c} N_\text{t}$ to $T \times N_\text{c} N_\text{t}$, then the single LSTM network \cite{grefflstm2017} is used to extract the temporal feature of $\textbf{V}$. 
% Note that the number of layers of the LSTM module equals to the length of time steps $T$. The output of the final LSTM module is the global temporal feature of  $\textbf{V}$.
% The computations involved in this LSTM module follow the typical design in \cite{grefflstm2017}
% \begin{subequations}
% \begin{equation}
%     o_t = \delta (W_{xo}x_t + h_{t-1}W_{ho} + b_o),
% \end{equation}
% \begin{equation}
%     c_t = f_t \odot c_{t-1} + i_t \odot \tilde{C}_t,
% \end{equation}
% \begin{equation}
%     y_t = o_t \odot h(c_t),
% \end{equation}
% \end{subequations}

% \noindent where $x_t$ and $y_t$ denote the input and the output of the LSTM module, $c_t$, $\tilde{C}_t$, and $h_t$ respectively denote time memory cell, candidate memory cell, and hidden state, $W$ and $b$ are respectively the weight parameters and the bias parameters, and $\delta$ is the activation function. Note that the number of layers of the LSTM module equals to the length of time steps $T$. The output of the final LSTM module is the global temporal feature of $T$-dimensional $\textbf{H}_\text{e}$.

\subsection{Feature Decoupling Decoder}
After receiving the codeword $\textbf{c}$ at BS, the feature decoupling decoder in Fig. 2 is designed to recover the CSI matrix, which consists of the spatial decompression layer, two convolutional neural network (CNN) blocks, and a normalization layer. The decoder inputs $\textbf{c}$ and outputs the recovered CSI matrix, $\widehat{\textbf{H}}_\text{c}$. The spatial decompression layer adopts a 1D convolutional layer with the filter size of $ 2 N_\text{c} N_\text{t} \times M \times 1$ to decompress the spatial features from $\textbf{c}$. The decompressed features are reshaped to an  matrix, denoted by $\textbf{W} \in \mathbb{R}^{T \times 2 \times N_\text{c} \times N_\text{t}}$, before forwarded to the CSI recovery module. The CSI recovery module contains two CNN blocks and a normalization layer. The parameters of these network layers are summarized in Table \uppercase\expandafter{\romannumeral1}. The value $0.3$ of LReLU function is a constraint parameter, which allows negative values of a matrix to be mapped to corresponding tiny values rather than simply all zeros.

\begin{table}[!t]  
\caption{Parameters of the 3D Convolutional Layers}
\vspace{-0.6em}
\centering
\setlength{\tabcolsep}{5mm}{
\begin{tabular}{crc}
\toprule
\multicolumn{3}{l}{{\bf Input}: $\textbf{W} \in \mathbb{R}^{T \times 2 \times N_\text{c} \times N_\text{t}}$, \ \ \ \ \ {\bf Output}: $\widehat{\textbf{H}}_\text{c} \in \mathbb{R}^{T \times 2 \times N_\text{c} \times N_\text{t}}$} \\  
\midrule
\multicolumn{3}{l}{\bf Convolutional layers}  \\ 
 {\bf Layers} & {\bf Filters/Stride/Padding} & {\bf Activation}\\
 (1) & $2\times 1 \times 7\times7/1/3$   & BN + $\text{LReLU}_{(0.3)}$   \\  
 (2) & $4\times 1 \times 5\times5/1/2$  & BN + $\text{LReLU}_{(0.3)}$   \\  
 (3) & $8\times 1 \times 5\times5/1/2$  & BN + $\text{LReLU}_{(0.3)}$   \\  
 (4) & $8\times 1 \times 3\times3/1/1$  & BN + $\text{LReLU}_{(0.3)}$   \\  
 (5) & $2\times 1 \times 1\times1/1/1$ & BN + $\text{LReLU}_{(0.3)}$   \\  
 (6) & $2\times 1 \times 3\times3/1/1$ & BN + $\text{Sigmoid}$ \\
% \midrule
% \multicolumn{3}{l}{{\bf Output}: $\widehat{\textbf{H}}_\text{c}$}\\
\bottomrule  
\end{tabular}}
\label{tb:label1}
\end{table}

\section{Experimental Results}
In this section, we verify the effectiveness of the proposed SD-CsiNet in terms of normalized mean-squared error (NMSE). Then we compare the number of parameters of SD-CsiNet with state-of-the-art methods. For practical application, we also compare the performance when quantized feedback codeword is further considered. Further, we conducts ablation experiments to explore the impact of each module of SD-CsiNet on CSI recovery.

\emph{1) Simulation Setting:} We generate $150,000$ channel samples through the COST $2100$ indoor channel model and the channel samples are split into $100,000$ for training sets, $30,000$ for validation sets and $20,000$ for testing sets. The BS with $N_\text{t} = 32$ antennas is deployed with a uniform linear array (ULA). The number of subcarriers of OFDM is $N_\text{s} = 1024$ and the first main-valued rows is $N_\text{c} = 32$. The time step of temporal CSI images is set to $T = 5$. The trainable weights and bias of all convolutional layers are initialized randomly. The simulation is carried out in Pytorch on a GTX3090 GPU.

% The warm up epoch is set to $T_\text{w} = 30$ and the total epoch is set to $T_\text{t} = 1000$. The batch size is set to 40. The initial lr is set to $\gamma_\text{min} = 2e-3$ and the final lr is set to $\gamma_\text{max} = 5e-5$. The simulation is carried out in Pytorch on a GTX3090 GPU.

\begin{figure*}[t]
\centering
\subfigure[$\sigma = \frac{1}{8}$.]{
\includegraphics[width = 7.3cm]{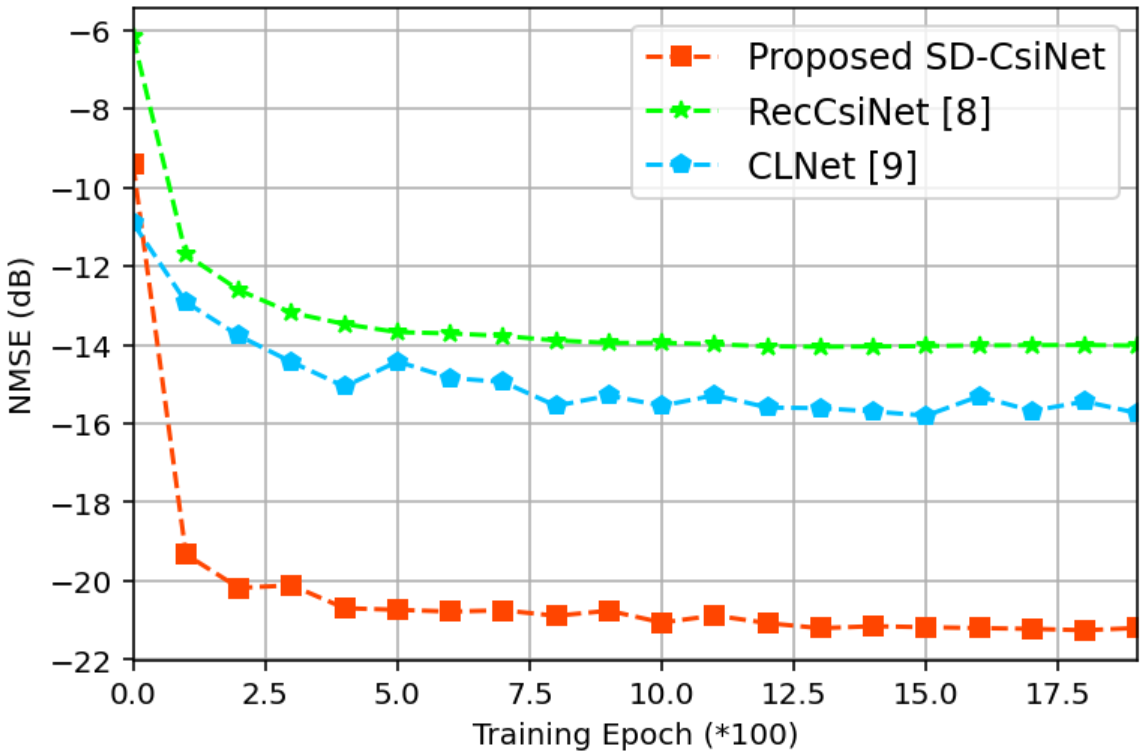}}
\subfigure[$\sigma = \frac{1}{16}$.]{
\includegraphics[width = 7.3cm]{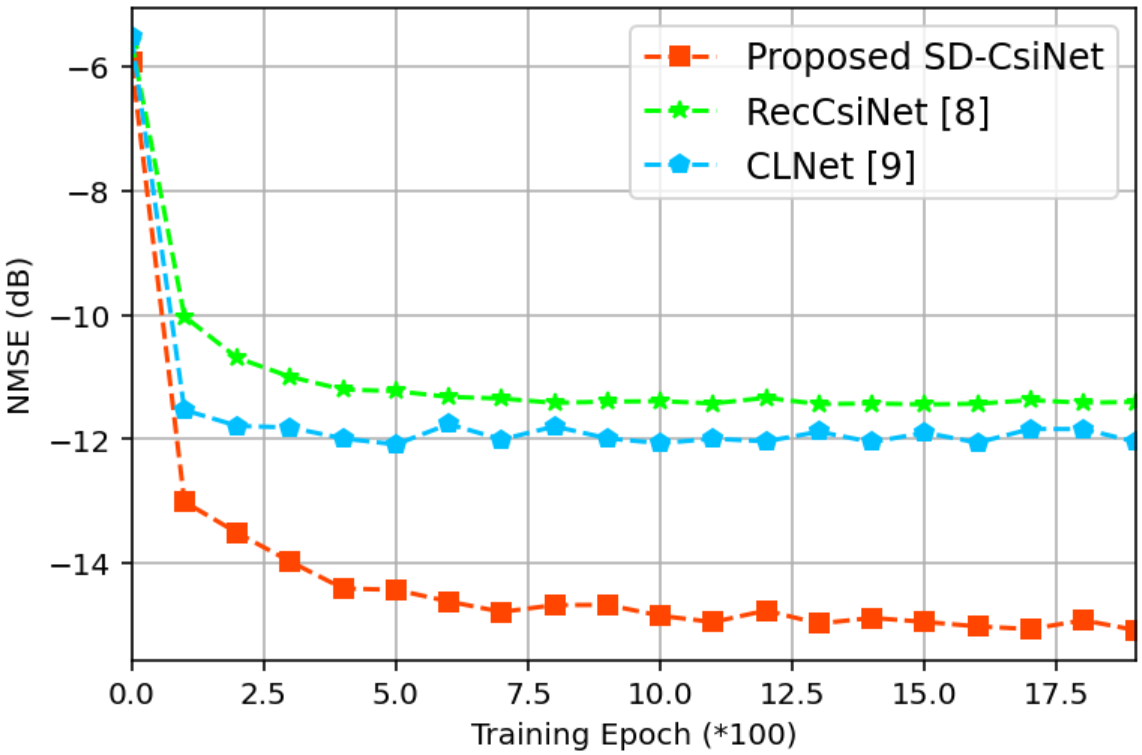}}
\subfigure[$\sigma = \frac{1}{32}$.]{
\includegraphics[width = 7.3cm]{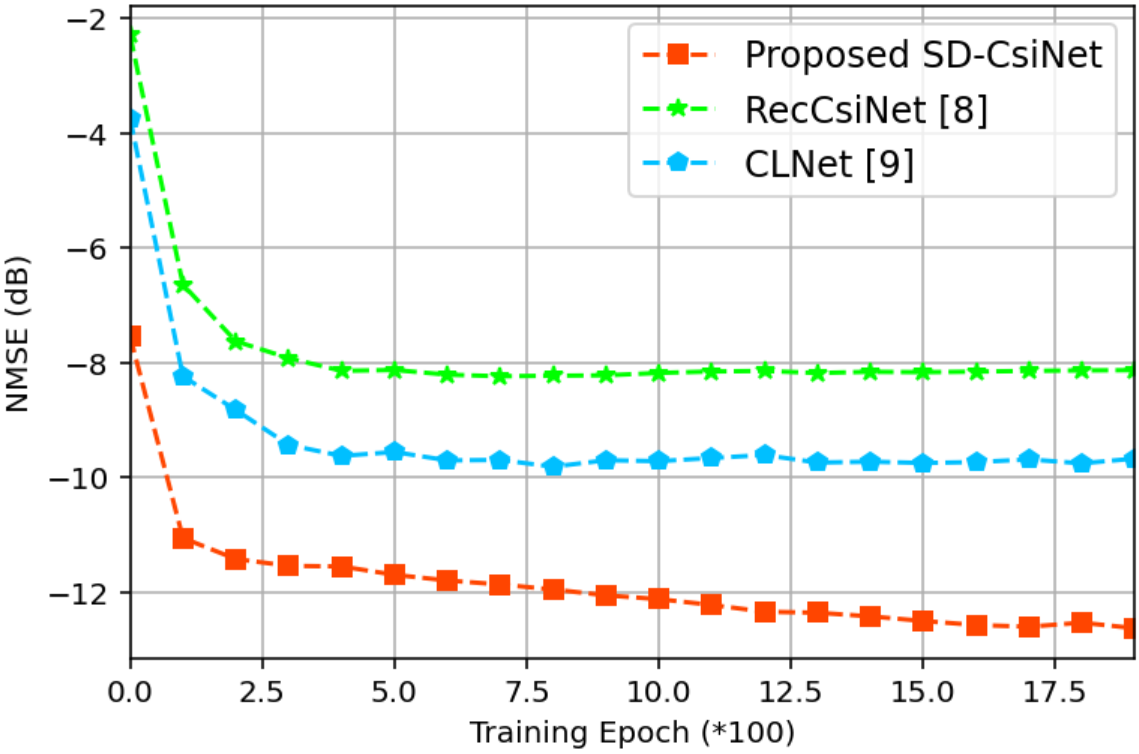}}
\subfigure[$\sigma = \frac{1}{64}$.]{
\includegraphics[width = 7.3cm]{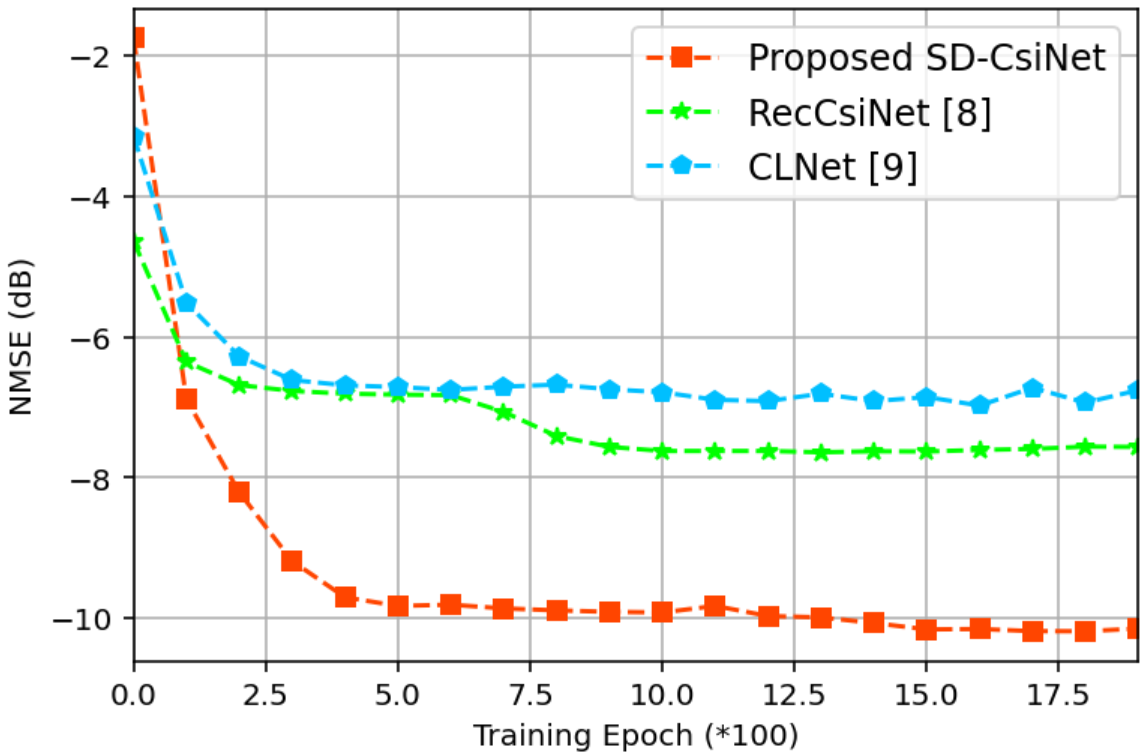}}
\caption{NMSE comparison under different compression ratios $\sigma$.}
\end{figure*}

\emph{2) NMSE Comparison:} To valid the performance of the proposed SD-CsiNet on the CSI recovery, we compare the NMSE with existing methods such as RecCsiNet \cite{lumimo2019} and CLNet \cite{jiclnet2021}. The NMSE is calculated by
\begin{equation}
    \text{NMSE} = E \left \{ \left \| \textbf{H}_\text{c} - \widehat{\textbf{H}}_\text{c} \right \|_2^2 / \left \| \textbf{H}_\text{c} \right \|_2^2 \right \},
\end{equation}
\noindent where notation $E \left \{ \cdot \right \}$ is the expectation operator. We carry out the experimental simulations with various compression ratios of $1/8$, $1/16$, $1/32$, and $1/64$. The comparison results are shown in Fig. 3 and we can observe that the proposed SD-CsiNet performs best under all compression ratios. The performance gain of NMSE for SD-CsiNet is similar under other compression ratios.

% As the compression ratio is $\sigma = 1/8$, the NMSE of SD-CsiNet is $-21.19$ dB. The proposed SD-CsiNet performs $7.17$ dB performance gain than RecCsiNet \cite{lumimo2019} and $5.45$ dB performance gain than CLNet \cite{jiclnet2021}.
% As the compression ratio is $\sigma = 1/16$, The proposed SD-CsiNet performs $3.68$ dB performance gain than RecCsiNet \cite{lumimo2019} and $3.03$ dB performance gain than CLNet \cite{jiclnet2021}.

\emph{3) Parameter Comparison:} As shown in Table II, we compare the network parameters of proposed SD-CsiNet with RecCsiNet \cite{lumimo2019} and CLNet \cite{jiclnet2021}. We can observe that the proposed SD-CsiNet has the least parameters under all compression ratios. The parameters of SD-CsiNet is $19$ M less than RecCsiNet on average. The parameters of SD-CsiNet is $30.59 \%$, $23.26 \%$, $29.27 \%$, and $28.57 \%$ than CLNet at the $\sigma$ of $1/8$, $1/16$, $1/32$, and $1/64$.

\begin{table}[t]
\caption{Network Complexity Comparison }
\vspace{-0.6em}
\centering
\setlength{\tabcolsep}{1mm}{
\begin{tabular}{@{}c|ccc|ccc|ccc|ccc@{}}
\toprule
\multirow{2}{*}{\diagbox [trim=l] {Methods} {$\sigma$}} & \multicolumn{3}{c|}{1/8}        & \multicolumn{3}{c|}{1/16}       & \multicolumn{3}{c|}{1/32}       & \multicolumn{3}{c}{1/64}        \\ \cmidrule(l){2-13} 
                            & UE    & BS     & Total          & UE    & BS     & Total          & UE    & BS     & Total          & UE    & BS     & Total          \\ \midrule
RecCsiNet \cite{lumimo2019}                  & 2.89M & 19.42M & 22.31M         & 1.38M & 18.11M & 19.49M         & 0.67M & 17.45M & 18.12M         & 0.33M & 17.13M & 17.46M         \\ \midrule
CLNet \cite{jiclnet2021}                      & 1.05M & 1.06M  & 2.11M          & 0.53M & 0.53M  & 1.06M          & 0.26M & 0.27M  & 0.53M          & 0.13M & 0.14M  & 0.27M          \\ \midrule
SD-CsiNet                   & 1.32M & 0.53M  & \textbf{1.85M} & 0.59M & 0.27M  & \textbf{0.86M} & 0.28M & 0.13M  & \textbf{0.41M} & 0.14M & 0.07M  & \textbf{0.21M} \\ \bottomrule
\end{tabular}}
\end{table}

\begin{table}[t]
\centering
\caption{NMSE Comparison With And Without Quantization}
\vspace{-0.6em}
\setlength{\tabcolsep}{1mm}{
\begin{tabular}{@{}ccccc@{}}
\toprule
Methods                        & \begin{tabular}[c]{@{}c@{}}Compression\\ ratio\end{tabular} & NMSE (dB)                             & NMSE-Q (dB)   & \begin{tabular}[c]{@{}c@{}}Online running\\ time \end{tabular}   \\ \midrule
\multicolumn{1}{c|}{RecCsiNet \cite{lumimo2019} } & \multicolumn{1}{c|}{\multirow{3}{*}{1/8}}                  & \multicolumn{1}{c|}{-14.02}           & -13.87    & 7.39 ms      \\
\multicolumn{1}{c|}{\text{CLNet \cite{jiclnet2021}}}   & \multicolumn{1}{c|}{}                                      & \multicolumn{1}{c|}{\text{-15.74}} & \text{-15.53}    & 5.13 ms    \\
\multicolumn{1}{c|}{\textbf{SD-CsiNet}}   & \multicolumn{1}{c|}{}                                      & \multicolumn{1}{c|}{\textbf{-21.19}} & \textbf{-20.76}    &  7.27 ms  \\ \midrule
\multicolumn{1}{c|}{RecCsiNet \cite{lumimo2019} } & \multicolumn{1}{c|}{\multirow{3}{*}{1/16}}                 & \multicolumn{1}{c|}{-11.40}           & -11.18    & 6.92 ms         \\
\multicolumn{1}{c|}{\text{CLNet \cite{jiclnet2021}}}   & \multicolumn{1}{c|}{}                                      & \multicolumn{1}{c|}{\text{-12.05}} & \text{-11.87}   & 5.08 ms  \\
\multicolumn{1}{c|}{\textbf{SD-CsiNet}}   & \multicolumn{1}{c|}{}                                      & \multicolumn{1}{c|}{\textbf{-15.08}} & \textbf{-14.84}  & 6.79 ms  \\ \bottomrule
\end{tabular}}
\end{table}

\emph{4) Comparison with Quantization Feedback:} In this paper, we train the SD-CsiNet without considering the quantization on the offline training stage. For the online deployment, we adopt the Lloyd-Max algorithm to quantize the codeword  $\textbf{c}$. The Lloyd-Max algorithm is a non-uniform quantization algorithm, which reduces the quantization interval when the probability density of the compressed codeword is large, and vice versa. In the simulation, each compressed codeword is quantized by $6$ bits. The quantization results with $\sigma = \left \{ 1/8,1/16 \right \}$ are compared in Table \uppercase\expandafter{\romannumeral3}. We observe that the proposed SD-CsiNet still outperforms RecCsiNet \cite{lumimo2019} and CLNet \cite{jiclnet2021} when the codeword is quantized. In Table III, we also compare the online computational complexity, i.e., online running time, of these methods. It is observed that all these  methods have similar online running time.

% In the practical application, it is necessary to quantize codeword values before sending back to the BS. In this paper,

\emph{5) Ablation Experiment:} To valid the effectiveness of each module of SD-CsiNet for CSI recovery, we conduct ablation experiment as shown in Table IV. The baseline is the CNN consists of naive convolutional layers and FC layers. The epoch of ablation experiment is set to $600$. From Table IV we can observe that each module of SD-CsiNet has the essential impact on high-accuracy CSI recovery.  Especially, the self-information image in the self-information domain explicitly represents the information contained in the CSI image. Therefore, the proposed SD-CsiNet can allocate more compressed resources to regions with large information in the self-information image, i.e., realize the “difference” allocation of compressed resources according to the information  amount, which results in the performance gain in CSI feedback and reconstruction.

\begin{table}[t]
\centering
\caption{Ablation Experiment for Proposed SD-CsiNet}
\vspace{-0.6em}
\begin{tabular}{@{}c|c|c|c|c@{}}
\toprule
CR   & Baseline & + LSTM & + SF Module & + LSTM + SF Module \\ \midrule
1/8  & -14.96 dB       & -15.11 dB     & -17.70 dB          & \textbf{-21.19 dB}          \\ \midrule
1/16 & -9.82 dB       & -12.13 dB     & -14.75 dB          & \textbf{-15.08 dB}          \\ \midrule
1/32 & -7.82 dB       & -10.77 dB     & -12.04 dB          & \textbf{-12.63 dB}          \\ \midrule
1/64 & -5.93 dB       & -8.29 dB     & -9.83 dB          & \textbf{-10.15 dB}          \\ \bottomrule
\end{tabular}
\end{table}

\section{Conclusion}
In this paper, we proposed a novel DL-based network, named SD-CsiNet, for the CSI compression and recovery. We exploited the self-information to represent the amount of information contained in the CSI. Then we extracted and coupled the spatial and temporal features of CSI from the perspective of self-information to achieve efficient compression. Further, the CSI matrix could be recovered accurately by decoupling these features. The experimental results showed that the proposed network could achieve the obvious NMSE performance gain compared to existing DL-based networks under all compression ratios. 

\bibliographystyle{IEEEtran}
\bibliography{SD-CsiNet.bib}


% \end{IEEEbiography}


%\newpage



%\vfill


\end{document}